\catcode`\@=11

\font\tenmsa=msam10
\font\sevenmsa=msam7
\font\fivemsa=msam5
\font\tenmsb=msbm10
\font\sevenmsb=msbm7
\font\fivemsb=msbm5
\newfam\msafam
\newfam\msbfam
\textfont\msafam=\tenmsa  \scriptfont\msafam=\sevenmsa
  \scriptscriptfont\msafam=\fivemsa
\textfont\msbfam=\tenmsb  \scriptfont\msbfam=\sevenmsb
  \scriptscriptfont\msbfam=\fivemsb

\def\hexnumber@#1{\ifnum#1<10 \number#1\else
 \ifnum#1=10 A\else\ifnum#1=11 B\else\ifnum#1=12 C\else
 \ifnum#1=13 D\else\ifnum#1=14 E\else\ifnum#1=15 F\fi\fi\fi\fi\fi\fi\fi}

\def\msa@{\hexnumber@\msafam}
\def\msb@{\hexnumber@\msbfam}
\mathchardef\boxdot="2\msa@00
\mathchardef\boxplus="2\msa@01
\mathchardef\boxtimes="2\msa@02
\mathchardef\square="0\msa@03
\mathchardef\blacksquare="0\msa@04
\mathchardef\centerdot="2\msa@05
\mathchardef\lozenge="0\msa@06
\mathchardef\blacklozenge="0\msa@07
\mathchardef\circlearrowright="3\msa@08
\mathchardef\circlearrowleft="3\msa@09
\mathchardef\rightleftharpoons="3\msa@0A
\mathchardef\leftrightharpoons="3\msa@0B
\mathchardef\boxminus="2\msa@0C
\mathchardef\Vdash="3\msa@0D
\mathchardef\Vvdash="3\msa@0E
\mathchardef\vDash="3\msa@0F
\mathchardef\twoheadrightarrow="3\msa@10
\mathchardef\twoheadleftarrow="3\msa@11
\mathchardef\leftleftarrows="3\msa@12
\mathchardef\rightrightarrows="3\msa@13
\mathchardef\upuparrows="3\msa@14
\mathchardef\downdownarrows="3\msa@15
\mathchardef\upharpoonright="3\msa@16

\mathchardef\downharpoonright="3\msa@17
\mathchardef\upharpoonleft="3\msa@18
\mathchardef\downharpoonleft="3\msa@19
\mathchardef\rightarrowtail="3\msa@1A
\mathchardef\leftarrowtail="3\msa@1B
\mathchardef\leftrightarrows="3\msa@1C
\mathchardef\rightleftarrows="3\msa@1D
\mathchardef\Lsh="3\msa@1E
\mathchardef\Rsh="3\msa@1F
\mathchardef\rightsquigarrow="3\msa@20
\mathchardef\leftrightsquigarrow="3\msa@21
\mathchardef\looparrowleft="3\msa@22
\mathchardef\looparrowright="3\msa@23
\mathchardef\circeq="3\msa@24
\mathchardef\succsim="3\msa@25
\mathchardef\gtrsim="3\msa@26
\mathchardef\gtrapprox="3\msa@27
\mathchardef\multimap="3\msa@28
\mathchardef\therefore="3\msa@29
\mathchardef\because="3\msa@2A
\mathchardef\doteqdot="3\msa@2B

\mathchardef\triangleq="3\msa@2C
\mathchardef\precsim="3\msa@2D
\mathchardef\lesssim="3\msa@2E
\mathchardef\lessapprox="3\msa@2F
\mathchardef\eqslantless="3\msa@30
\mathchardef\eqslantgtr="3\msa@31
\mathchardef\curlyeqprec="3\msa@32
\mathchardef\curlyeqsucc="3\msa@33
\mathchardef\preccurlyeq="3\msa@34
\mathchardef\leqq="3\msa@35
\mathchardef\leqslant="3\msa@36
\mathchardef\lessgtr="3\msa@37
\mathchardef\backprime="0\msa@38
\mathchardef\risingdotseq="3\msa@3A
\mathchardef\fallingdotseq="3\msa@3B
\mathchardef\succcurlyeq="3\msa@3C
\mathchardef\geqq="3\msa@3D
\mathchardef\geqslant="3\msa@3E
\mathchardef\gtrless="3\msa@3F
\mathchardef\sqsubset="3\msa@40
\mathchardef\sqsupset="3\msa@41
\mathchardef\trianglerighteq="3\msa@44
\mathchardef\trianglelefteq="3\msa@45
\mathchardef\bigstar="0\msa@46
\mathchardef\between="3\msa@47
\mathchardef\blacktriangledown="0\msa@48
\mathchardef\blacktriangleright="3\msa@49
\mathchardef\blacktriangleleft="3\msa@4A
\mathchardef\blacktriangle="0\msa@4E
\mathchardef\triangledown="0\msa@4F
\mathchardef\eqcirc="3\msa@50
\mathchardef\lesseqgtr="3\msa@51
\mathchardef\gtreqless="3\msa@52
\mathchardef\lesseqqgtr="3\msa@53
\mathchardef\gtreqqless="3\msa@54
\mathchardef\Rrightarrow="3\msa@56
\mathchardef\Lleftarrow="3\msa@57
\mathchardef\veebar="2\msa@59
\mathchardef\barwedge="2\msa@5A
\mathchardef\doublebarwedge="2\msa@5B
\mathchardef\angle="0\msa@5C
\mathchardef\measuredangle="0\msa@5D
\mathchardef\sphericalangle="0\msa@5E
\mathchardef\varpropto="3\msa@5F
\mathchardef\smallsmile="3\msa@60
\mathchardef\smallfrown="3\msa@61
\mathchardef\Subset="3\msa@62
\mathchardef\Supset="3\msa@63
\mathchardef\Cup="2\msa@64

\mathchardef\Cap="2\msa@65

\mathchardef\curlywedge="2\msa@66
\mathchardef\curlyvee="2\msa@67
\mathchardef\leftthreetimes="2\msa@68
\mathchardef\rightthreetimes="2\msa@69
\mathchardef\subseteqq="3\msa@6A
\mathchardef\supseteqq="3\msa@6B
\mathchardef\bumpeq="3\msa@6C
\mathchardef\Bumpeq="3\msa@6D
\mathchardef\lll="3\msa@6E

\mathchardef\ggg="3\msa@6F

\mathchardef\circledS="0\msa@73
\mathchardef\pitchfork="3\msa@74
\mathchardef\dotplus="2\msa@75
\mathchardef\backsim="3\msa@76
\mathchardef\backsimeq="3\msa@77
\mathchardef\complement="0\msa@7B
\mathchardef\intercal="2\msa@7C
\mathchardef\circledcirc="2\msa@7D
\mathchardef\circledast="2\msa@7E
\mathchardef\circleddash="2\msa@7F
\def\ulcorner{\delimiter"4\msa@70\msa@70 }
\def\urcorner{\delimiter"5\msa@71\msa@71 }
\def\llcorner{\delimiter"4\msa@78\msa@78 }
\def\lrcorner{\delimiter"5\msa@79\msa@79 }
\def\yen{\mathhexbox\msa@55 }
\def\checkmark{\mathhexbox\msa@58 }
\def\circledR{\mathhexbox\msa@72 }
\def\maltese{\mathhexbox\msa@7A }
\mathchardef\lvertneqq="3\msb@00
\mathchardef\gvertneqq="3\msb@01
\mathchardef\nleq="3\msb@02
\mathchardef\ngeq="3\msb@03
\mathchardef\nless="3\msb@04
\mathchardef\ngtr="3\msb@05
\mathchardef\nprec="3\msb@06
\mathchardef\nsucc="3\msb@07
\mathchardef\lneqq="3\msb@08
\mathchardef\gneqq="3\msb@09
\mathchardef\nleqslant="3\msb@0A
\mathchardef\ngeqslant="3\msb@0B
\mathchardef\lneq="3\msb@0C
\mathchardef\gneq="3\msb@0D
\mathchardef\npreceq="3\msb@0E
\mathchardef\nsucceq="3\msb@0F
\mathchardef\precnsim="3\msb@10
\mathchardef\succnsim="3\msb@11
\mathchardef\lnsim="3\msb@12
\mathchardef\gnsim="3\msb@13
\mathchardef\nleqq="3\msb@14
\mathchardef\ngeqq="3\msb@15
\mathchardef\precneqq="3\msb@16
\mathchardef\succneqq="3\msb@17
\mathchardef\precnapprox="3\msb@18
\mathchardef\succnapprox="3\msb@19
\mathchardef\lnapprox="3\msb@1A
\mathchardef\gnapprox="3\msb@1B
\mathchardef\nsim="3\msb@1C
\mathchardef\napprox="3\msb@1D
\mathchardef\nsubseteqq="3\msb@22
\mathchardef\nsupseteqq="3\msb@23
\mathchardef\subsetneqq="3\msb@24
\mathchardef\supsetneqq="3\msb@25
\mathchardef\subsetneq="3\msb@28
\mathchardef\supsetneq="3\msb@29
\mathchardef\nsubseteq="3\msb@2A
\mathchardef\nsupseteq="3\msb@2B
\mathchardef\nparallel="3\msb@2C
\mathchardef\nmid="3\msb@2D
\mathchardef\nshortmid="3\msb@2E
\mathchardef\nshortparallel="3\msb@2F
\mathchardef\nvdash="3\msb@30
\mathchardef\nVdash="3\msb@31
\mathchardef\nvDash="3\msb@32
\mathchardef\nVDash="3\msb@33
\mathchardef\ntrianglerighteq="3\msb@34
\mathchardef\ntrianglelefteq="3\msb@35
\mathchardef\ntriangleleft="3\msb@36
\mathchardef\ntriangleright="3\msb@37
\mathchardef\nleftarrow="3\msb@38
\mathchardef\nrightarrow="3\msb@39
\mathchardef\nLeftarrow="3\msb@3A
\mathchardef\nRightarrow="3\msb@3B
\mathchardef\nLeftrightarrow="3\msb@3C
\mathchardef\nleftrightarrow="3\msb@3D
\mathchardef\divideontimes="2\msb@3E
\mathchardef\varnothing="0\msb@3F
\mathchardef\nexists="0\msb@40
\mathchardef\mho="0\msb@66
\mathchardef\thorn="0\msb@67
\mathchardef\beth="0\msb@69
\mathchardef\gimel="0\msb@6A
\mathchardef\daleth="0\msb@6B
\mathchardef\lessdot="3\msb@6C
\mathchardef\gtrdot="3\msb@6D
\mathchardef\ltimes="2\msb@6E
\mathchardef\rtimes="2\msb@6F
\mathchardef\shortmid="3\msb@70
\mathchardef\shortparallel="3\msb@71
\mathchardef\smallsetminus="2\msb@72
\mathchardef\thicksim="3\msb@73
\mathchardef\thickapprox="3\msb@74
\mathchardef\approxeq="3\msb@75
\mathchardef\succapprox="3\msb@76
\mathchardef\precapprox="3\msb@77
\mathchardef\curvearrowleft="3\msb@78
\mathchardef\curvearrowright="3\msb@79
\mathchardef\digamma="0\msb@7A
\mathchardef\varkappa="0\msb@7B
\mathchardef\hslash="0\msb@7D
\mathchardef\hbar="0\msb@7E
\mathchardef\backepsilon="3\msb@7F
\def\Bbb{\ifmmode\let\next\Bbb@\else
 \def\next{\errmessage{Use \string\Bbb\space only in math mode}}\fi\next}
\def\Bbb@#1{{\Bbb@@{#1}}}
\def\Bbb@@#1{\fam\msbfam#1}

\catcode`\@=\active

\def\CR{\hbox{{$\cal R$}}}

\def\R{{\Bbb R}}
\def\C{{\Bbb C}}

\def\vect{{\bf t}}\def\vecs{{\bf s}}
\def\vecu{{\bf u}}\def\vecx{{\bf x}}\def\vecp{{\bf p}}

\def\vecz{{\bf z}}

\def\<{\langle}
\def\>{\rangle}

\def\lform{\hbox{$\sqcup$}\llap{\hbox{$\sqcap$}}}

\def\h{{{1\over2}}}
\def\eps{{\epsilon}}

\def\rcross{{\triangleright\!\!\!<}}

\def\dcross{{\bowtie}}
\def\codcross{{\blacktriangleright\!\!\blacktriangleleft}}

\def\nosum{{}}
\def\tens{\mathop{\otimes}}
\def\la{{\triangleright}}\def\ra{{\triangleleft}}
\def\isom{{\cong}}

\def\Ad{{\rm Ad}}

\def\id{{\rm id}}

\def\proof{\goodbreak\noindent{\bf Proof\quad}}

\def\endproof{{\ $\lform$}\bigskip }

\def\und#1{{\underline {#1}}}

\def\o{{}_{\scriptscriptstyle(1)}}
\def\t{{}_{\scriptscriptstyle(2)}}
\def\th{{}_{\scriptscriptstyle(3)}}
\def\fo{{}_{\scriptscriptstyle(4)}}

\def\uo{{{}^{\scriptscriptstyle(1)}}}
\def\ut{{{}^{\scriptscriptstyle(2)}}}

\def\note#1{}

\def\equad{\kern -1.7em}

\def\eqn#1#2{\begin{equation}#2\label{#1}\end{equation}}
\def\cmath#1{\[\begin{array}{c} #1 \end{array}\]}

\def\ceqn#1#2{\begin{equation}
\label{#1}\begin{array}{c}#2\end{array}\end{equation}}
\def\align#1{\begin{eqnarray*}#1\end{eqnarray*}}

\def\dila{{\varsigma}} 

\documentstyle[12pt]{article}
\newtheorem{lemma}{Lemma}[section]
\newtheorem{propos}[lemma]{Proposition}

\begin{document}\baselineskip 19pt

{\ }\hskip 4.7in DAMTP/94-03 
\vspace{.2in}

\begin{center} {\LARGE {$q$}-EUCLIDEAN SPACE AND QUANTUM GROUP WICK ROTATION BY
TWISTING}
\\ \baselineskip 13pt{\ }
{\ }\\ S. Majid\footnote{Royal Society University Research Fellow and Fellow of
Pembroke College, Cambridge} \note{\& U. Meyer}\\ {\ }\\
Department of Applied Mathematics \& Theoretical Physics\\ University of
Cambridge, Cambridge CB3 9EW, U.K.
\end{center}

\begin{center}
January, 1994\end{center}
\vspace{10pt}
\begin{quote}\baselineskip 13pt
\noindent{\bf Abstract} We study the quantum matrix algebra
$R_{21}\vecx_1\vecx_2=\vecx_2\vecx_1 R$ and for the standard $2\times 2$ case
propose it for the co-ordinates of $q$-deformed Euclidean space. The algebra in
this simplest case is isomorphic to the usual quantum matrices $M_q(2)$ but in
a form which is naturally covariant under the Euclidean rotations $SU_q(2)\tens
SU_q(2)$. We also introduce a quantum Wick rotation that twists this system
precisely into the approach to $q$-Minkowski space based on braided-matrices
and their associated spinorial $q$-Lorentz group.  \end{quote}
\baselineskip 21pt

\section{Introduction}

The problem of constructing a $q$-deformed or non-commutative analogue of
spacetime co-ordinates has been extensively studied in recent years. One
established approach is based on the idea that $q$-Minkowski space should be
modelled by $2\times 2$ braided matrices\cite{Ma:exa}\cite{Ma:mec}, which
approach turned out to be compatible also with the approaches of
\cite{CWSSW:ten}\cite{CWSSW:lor}\cite{OSWZ:def} based on other ideas of spinor
decomposition and with \cite{PodWor:def} based on the idea that the $q$-Lorentz
group should be the quantum double. The braided matrix approach has been
developed in detail by U. Meyer\cite{Mey:new} who introduced a (braided)
coaddition law for $q$-Minkowski space 4-vectors as well as clarifying
covariance and other properties. The connection with the quantum double point
of view was established in \cite[Sec.4]{Ma:poi} while the connection with the
spinorial approach was established in \cite{MaMey:bra}. The identification of
the corresponding $q$-Lorentz group quantum enveloping algebra as a `twisted
square'\cite{ResSem:mat} was introduced by the author in \cite[Sec. 4]{Ma:poi}.

In this paper we supplement this line of development with a fully compatible
proposal for $q$-Euclidean space. It is obvious from the above context that
$SU_q(2)\tens SU_q(2)$ should be the corresponding rotation quantum group in
the Euclidean case within a $2\times 2$ matrix approach, which will be our
point of view also. Not known however, is what should play the role of
$q$-Euclidean space itself. We need once again a matrix algebra, but neither
the braided-matrices (which are naturally hermitian) nor $2\times 2$ quantum
matrices in their usual form (which have the right $*$-structure but are not
properly covariant under the obvious matrix action) will do. Rather, we give a
general construction in which we start with a copy of the usual quantum
matrices $A(R)$ with generators $\vect$ and define on their linear space a new
product. The new algebra $\bar A(R)$ has generators $\vecx=\vect$ but different
products according to
\[ \vecx_1\cdot\vecx_2=\lambda_0R\vect_1\vect_2 \]
giving the commutation relations in the abstract. This is given in Section~3 of
the paper.

Our general strategy here is to introduce a new technique for
`covariantisation' by twisting of module algebras and comodule algebras. It is
well-known that a 2-cocycle on a quantum group allows one to `twist' the
quantum group\cite{Dri:qua} to a new one by conjugation. Less well known is
that this same cocycle can also be used to twist anything on which the quantum
group acts. This general theory is recalled first in Section~2.

In Section~4, we study the properties of this algebra $\bar A(R)$ in the
standard $2\times 2$ case. Thus we apply our construction to the usual quantum
matrices $M_q(2)$ of \cite{Dri}\cite{FRT:lie} and obtain our algebra $\bar
M_q(2)$ which we call we call $\R_q^4$ or $q$-Euclidean space. It turns out to
be isomorphic as an algebra to the usual $M_q(2)$, but not as a $*$-algebra or
coalgebra.

Finally, in Section~5 we show that out twisting theory for algebras on which
quantum groups act, can be used once more to twist our $q$-Euclidean algebra
into exactly the above-mentioned $q$-Minkowski space algebra. The twisting
cocycle also takes $SU_q(2)\tens SU_q(2)$ into the $q$-Lorentz group in the
correct way. In this way we have a precise notion of quantum Wick-rotation.

Note that in our original work \cite{Ma:exa}, we obtained the braided-matrices
algebra by a novel procedure of `transmutation' from the usual quantum
matrices. Thus too involves a new algebra $B(R)$ built on $A(R)$ with
generators $\vecu=\vect$ but a new `transmuted' product\cite{Ma:lin}.
At a mathematical level, our new result is to express such transmutation of a
quantum group into a braided one as a two-step twisting. The first twisting
gives $q$-Euclidean space and the second twisting gives  $q$-Minkowski space.

\section{Preliminaries: Twisting of Comodule Algebras}

We begin with some general constructions for quantum groups, needed in
Section~3 and again in Section~5. By quantum group we mean a quasitriangular
Hopf algebra $H$ of enveloping algebra type (with universal R-matrix
$\CR$\cite{Dri}). Later on we will cover also the case of a
dual-quasitriangular Hopf algebra $A$ (of function algebra type) where this
time $\CR:A\tens A\to \C$ is the `universal R-matrix functional'\cite{Ma:bg}.

Here $H$ is assumed to have a coproduct $\Delta$, counit $\eps$, and an
antipode $S$ etc. A 2-cocycle in $H$ is an invertible element $\chi\in H\tens
H$ such that
\ceqn{2-cocycle}{ \chi_{23}(\id\tens\Delta)
\chi=\chi_{12}(\Delta\tens\id)\chi,\quad (\eps\tens\id)\chi=1.}
In this case an easy application of ideas of Drinfeld in \cite{Dri:qua} tells
us that the same algebra with
 \eqn{twiH}{ \Delta_\chi =\chi(\Delta\ )\chi^{-1},\quad \CR_\chi=\chi_{21}\CR
\chi^{-1},\quad S_\chi=U(S\ )U^{-1}}
remains a quasitriangular Hopf algebra, which we denote $H_\chi$. Here
$U=\nosum \chi\uo (S \chi\ut  )$. This is well-known by now. One of the first
places where is was explicitly studied in this Hopf algebra setting is
\cite{GurMa:bra}.

To this theory, we need to add the following less well-known observation: If
$B$ is an algebra on which $H$ acts covariantly in the sense
\eqn{H-cov}{h\la(ab)=(h\o\la a)(h\t\la b),\quad h\la 1=\eps(h)1}
then it easy to see from the 2-cocycle condition that
\eqn{twistmodalg}{ a\cdot_\chi b=\cdot\chi^{-1}\la (a\tens b)}
defines a new associative algebra which we call $B_\chi$, and that this is
covariant in the sense of (\ref{H-cov}) under $H_\chi$. So the covariant system
consisting of $(H,B)$ is completely `gauge transformed' to $(H_\chi,B_\chi)$.
To see the covariance of the new system, one needs only to check
\align{&&\equad h\la (a\cdot_\chi b)=h\la\cdot (\chi^{-1}\la (a\tens
b)=\cdot\left((\Delta h)\chi\la (a\tens
b)\right)=\cdot\left(\chi^{-1}(\Delta\chi h)\la (a\tens
b)\right)=\cdot_\chi\left((\Delta_\chi h)\la(a\tens b)\right)}
as required.

In practice, it is convenient for us not to work with quantum enveloping
algebras but with quantum function algebras. In this case we need the dual
theory to that above. Thus a 2-cocycle on $A$ is $\chi:A\tens A\to \C$ such
that
\eqn{2-cocycle-on}{\chi(b\o\tens c\o)\chi(a\tens b\t c\t)=\chi(a\o\tens
b\o)\chi(a\t b\t\tens c)}
for all $a,b,c$. This exactly generalises a group 2-cocycle. We assume $\chi$
is convolution-invertible in the sense that there is $\chi^{-1}$ which is
inverse relative to $\chi$ to $\Delta$ in a standard way. We also assume that
$\chi(a\tens 1)=\chi(1\tens a)=\eps(a)$. Then the dual of Drinfeld's twisting
is
that
\eqn{cotwistprod}{ a\cdot_\chi b=\sum \chi(a\o\tens b\o)a\t b\t
\chi^{-1}(a\th\tens b\th)}
\eqn{cotwistR}{\CR_{\chi}(a\tens b)=\sum \chi(b\o\tens a\o)\CR(a\t\tens b\t)
\chi^{-1}(a\th\tens b\th)}
\eqn{cotwistant}{S_\chi(a)=\sum U(a\o)Sa\t U^{-1}(a\th),\quad U(a)=\sum
\chi(a\o\tens Sa\t).}
is also a dual-quasitriangular Hopf algebra $A_\chi$ say. Likewise, if $B$ is
an algebra coacted upon by $A$ in a covariant way (a comodule-algebra) then
\eqn{cotwistcomod}{ c\cdot_\chi d= \chi^{-1}(c\o\tens d\o) c\t d\t}
is a new algebra $B_\chi$ coacted upon covariantly by $A_\chi$. The covariance
condition is that the coaction $\beta$ should be an algebra homomorphism. Thus
the covariant system $(A,B)$ is completely `gauge transformed' to the new
system $(A_\chi,B_\chi)$ with the same coaction $\beta$. An introduction to the
abstract terminology here is in \cite{Ma:qua}, while a recent paper on the
theory of twisting is in \cite{Ma:clau}

\section{Construction of quantum algebras $\bar A(R)$}

Every Hopf algebra $A$ coacts on itself by the coproduct $\Delta$ both from the
left and from the right. If we are interested only in coactions from the right,
we can convert the left coaction to a right one as follows. Firstly
\eqn{betaL}{ \beta(a)=a\t\tens Sa\o}
is a right coaction, but it is not covariant (a comodule algebra) under $A$ but
rather under $A^{\rm op}$ because the antipode $S$ is an anti-homomorphism.

\begin{propos} Consider $A$ as a comodule algebra under $A^{\rm op}$ by
(\ref{betaL}). Consider $\chi=\CR^{-1}$ as a 2-cocycle on $A^{\rm op}$. Then
the twisted covariant system to $(A^{\rm op},A)$ is $(A,\bar A)$ where $\bar A$
is $A$ equipped with the new product
\[ c\bar\cdot d=\CR(c\o\tens d\o)c\t d\t \]
\end{propos}
\proof Here $\CR^{-1}$ is a dual-quasitriangular structure on $A^{\rm op}$
because $\CR$ is one on $A$. Twisting by it gives the product of $A^{\rm
op}_\chi$ from (\ref{cotwistprod}) as $ a\cdot_\chi b=\sum \CR^{-1}(a\o\tens
b\o)a\t b\t \CR(a\th\tens b\th)=ab$ since a dual-quasitriangular Hopf algebra
is commutative up to conjugation by $\CR$ (it obeys axioms dual to those of
Drinfeld in \cite{Dri}), cf \cite{Kem:mul} for twisting a quantum enveloping
algebra to its opposite coproduct. On the other hand, the copy of $A$ on which
$A^{\rm op}$ acted is also twisted and from (\ref{cotwistcomod}) its becomes
$\bar A$ as stated. The theory in the preliminaries ensures that $\bar A$ is an
associative algebra and that $A$ coacts on it by (\ref{betaL}). \endproof

Note that it is not really necessary here for the $A$ that is acted upon to be
a quantum group in that the formulae for $\bar A$ works also at the bialgebra
level. The acting copy needs to be a quantum group for (\ref{betaL}) to be
defined. We consider now the simplest matrix quantum group case. Thus consider
the bialgebras $A(R)$ as in \cite{FRT:lie}\cite{Dri} for $R$ a solution of the
QYBE (Quantum Yang-Baxter Equations). We suppose that $A(R)$ also has a
quotient $A$ which is a dual-quasitriangular Hopf algebra with antipode. The
dual quasitriangular structure is the one introduced by the author in an
equivalent form in \cite{Ma:qua} and is defined by
$\CR(\vect_1\tens\vect_2)=\lambda_0 R$ on the generators, extended uniquely to
the whole algebra. Here $\lambda_0$ is a normalisation parameter and can be set
equal to $1$ if $R$ is given in the quantum group normalisation
as explained in \cite{Ma:lin}.

\begin{propos} Let $A(R)$ with matrix generator $\vecz$ be coacted from the
right by $A^{\rm op}$ as
\[ z^i{}_j\mapsto z^a{}_j\tens S\bar t^i{}_a,\quad {\rm i.e.}\quad \vecz\to
\bar\vect^{-1}\vecz \]
where $\bar\vect$ is the matrix generator of $A^{\rm op}$. Then this covariant
system $(A^{\rm op},A(R))$ twists under $\chi=\CR^{-1}$ to the covariant system
$(A, \bar A(R))$ where $\bar A(R)$ has generators $\vecx$ and relations and
covariance
\[ R_{21}\vecx_1\vecx_2=\vecx_2\vecx_1 R,\quad \vecx\to \vect^{-1}\vecx\]
under the usual matrix generator $\vect$ of $A$.
\end{propos}
\proof We are in the setting of Section~2 with the initial system being the
quantum group $A^{\rm op}$ coacting on $B=A(R)$ as stated. We take twisting
cocycle $\chi=\CR^{-1}$ and twisting by this, the acting quantum group $A^{\rm
op}$ and the acted-upon $A(R)$ have new products
\cmath{ \bar\vect_1\cdot_\chi \bar\vect_2=R^{-1}\bar\vect_1\bar\vect_2
R=\bar\vect_2\bar\vect_1=R_{21}\bar\vect_2\cdot_\chi \bar\vect_1 R_{21}^{-1}\\
\vecz_1\cdot_\chi\vecz_2=\lambda_0 R\vecz_1\vecz_2=\vecz_2\vecz_1\lambda_0
R=R_{21}^{-1}\vecz_2\cdot_\chi\vecz_1 R.}
We see that $A^{\rm op}$ with its generators $\bar\vect$ becomes twisted back
to the usual $A$ with matrix generators $\vect$ say. Meanwhile, the relations
for $\bar A(R)$ come out as stated with $\vecx$ denoting $\vecz$ with the new
product. \endproof

Before studying this new algebra $\bar A(R)$ in detail, we consider now in the
same spirit our general Hopf algebra $A$ as an $A^{\rm op}\tens A$-comodule
algebra, where we view the left coaction as a right coaction of $A^{\rm op}$ as
in (\ref{betaL}) and at the same time make a usual right coaction via $\Delta$.
So the combined coaction is
\eqn{betaLR}{ \beta(a)=a\t\tens Sa\o\tens a\th }
We take on the Hopf algebra $A^{\rm op}\tens A$ the 2-cocycle
\eqn{chiLR}{ \chi((a\tens b)\tens (c\tens d))=\CR^{-1}(a\tens c)\eps(b)\eps(d)}
which is $\CR^{-1}$ on $A^{\rm op}$ and trivial on $A$.

\begin{propos} The algebra $\bar A$ of Proposition~3.1 is covariant under
$A\tens A$ coacting as in (\ref{betaLR}).
\end{propos}
\proof The twisting of the covariant system $(A^{\rm op}\tens A, A)$ coacting
as in (\ref{betaLR}) by the 2-cocycle in (\ref{chiLR}) is $(A\tens A,\bar A)$
where the computation is strictly as in the proof of Proposition~3.1. The
cocycle is trivial in the $A$ part of $A^{\rm op}\tens A$ so no further change
is caused by this additional part of the coaction (\ref{betaLR}). Thus $\bar A$
is as before. Also in the same way, twisting changes $A^{\rm op}\tens A$ to
$A\tens A$. \endproof

In the quantum matrix example, we learn that $\bar A(R)$ is covariant under
\eqn{matbetaLR}{x^i{}_j\mapsto x^a{}_b\tens (Ss^i{}_a)t^b{}_j,\quad {\rm i.e.
}\quad \vecx\mapsto \vecs^{-1}\vecx\vect,\quad [\vecs,\vect]=0}
where $A\tens A$ has matrix generators $\vecs,\vect$ for its two factors.

Another general feature of interest to us concerns $*$-structures on our Hopf
algebra or bialgebra $A$. A $*$-bialgebra means that the algebra is a
$*$-algebra and $(*\tens *)\circ\Delta=\Delta\circ *$. We say that a
dual-quasitriangular structure on it is of real-type if $\overline{\CR(a\tens
b)}=\CR(b^*\tens a^*)$\cite{Ma:poi}\cite{Ma:mec}. We have

\begin{propos} If $A$ is a $*$-bialgebra and $\CR$ real then $\bar A$ is a
$*$-algebra too with the same  $*$-operation\note{, and the coactions
(\ref{betaL}) and (\ref{betaLR}) of $A,A\tens A$ are $*$-algebra homomorphism.}
\end{propos}
\proof We have that $\bar A$ is a $*$-algebra as
\[ (a\bar\cdot b)^*= \overline{\CR(a\o\tens b\o)}b^*\t a^*\t=b^*\bar\cdot a^*\]
precisely because $\CR$ is of real type.\endproof

Finally, we remark that there is an operator $\psi:\bar A\tens \bar A\to \bar
A\tens \bar A$ defined by
\eqn{psi}{ \psi(a\tens b)=b\o\tens  \CR^{-1}(a\o\tens b\t)a\t}
such that
\eqn{psiprod}{ (\bar\cdot\tens\id)\circ\psi_{23}\circ\psi_{12} =\psi\circ
(\id\tens\bar\cdot),\quad (\id\tens\bar\cdot)\circ\psi_{12}\circ\psi_{23}
=\psi\circ (\bar\cdot\tens\id)}
\eqn{psidelta}{ \Delta (a\bar\cdot b)=a\o\bar\cdot \psi(a\t\tens b\o)\bar\cdot
b\t}
where we apply $\psi$ as shown and multiply its output from the left by $a\o$
and from the right by $b\t$, using the product of $\bar A$. This is an
elementary computation from the definitions above and the axioms of a
dual-quasitriangular structure. It means that $(\bar A,\Delta,\psi)$ (the new
product and the original coproduct) is something like a
braided-group\cite{Ma:bg} where the tensor products of $\bar A$ must be treated
with non-commuting statistics according to
\eqn{psibtens}{ (a\tens b)(c\tens d)=a\bar\cdot\psi(b\tens c)\bar\cdot d.}
Such a product law for tensor products of $\bar A$ is characteristic of a
braided tensor product\cite{Ma:exa}\cite{Ma:bra} but does not in fact require
$\psi$ to be a braiding. We need only (\ref{psiprod}) in order for this to give
an associative algebra on $\bar A\tens\bar A$. Then (\ref{psiprod}) says that
$\Delta$ is an algebra homomorphism with respect to this product. In general,
$\psi$ in (\ref{psi})  does {\em not} obey the QYBE nor commute in the required
way with $\Delta$ for a true braided-group (or braided-Hopf algebra) so we have
something slightly weaker.

In the quantum matrix example it means that we have an operator
\eqn{psimat}{ \psi(\vecx_1\tens\vecx_2)=\vecx_2\tens
\lambda_0^{-1}R^{-1}\vecx_1,\quad {\rm i.e.} \quad \vecx'_1\vecx_2=\vecx_2
\lambda_0^{-1}R^{-1}\vecx_1'}
which extends to products via (\ref{psiprod}) and defines the tensor product
algebra for two copies $\vecx,\vecx'$ say with relations as stated. The
coproduct $\Delta \vecx=\vecx\tens\vecx$ is an algebra homomorphism to this
algebra, i.e.,   $\vecx''=\vecx\vecx'$ obeys the relations of $\bar A(R)$ if
$\vecx,\vecx'$ do and have the cross relations shown. On the other hand, $\psi$
does {\em not} in general obey the QYBE (and moreover is not covariant under
(\ref{matbetaLR})). This weaker structure can be viewed as falling in between
the usual quantum matrices bialgebra $A(R)$ with coproduct
$\Delta\vect=\vect\tens\vect$ and the braided matrices $B(R)$ with coproduct
$\Delta\vecu=\vecu\tens\vecu$ into which it will become under the q-Wick
rotation in Section~5.

More important for physics is not the multiplication law, but the addition law.
Here we do have a braided-group whenever $R$ is Hecke with eigenvalues
$q,-q^{-1}$ say for the associated braiding operator. The additive braiding and
braid-statistics are then
\eqn{Psimat}{ \Psi(\vecx_1\tens\vecx_2)=R(\vecx_2\tens \vecx_1)R,\quad {\rm
i.e.}\quad \vecx_1'\vecx_2=R\vecx_2\vecx_1' R}
extending to products as a braiding and corresponding to the additive braid
statistics shown. Thus $\vecx''=\vecx+\vecx'$ obeys the relations of $\bar
A(R)$ if $\vecx,\vecx'$ do and have the mutual relations in (\ref{Psimat}).
This is proven following exactly the same steps as in
\cite{Ma:poi}\cite{Ma:add} for the braided-addition law of $A(R)$, to which we
refer the reader for details.

\begin{propos} The addition law on $\bar A(R)$ with the additive braiding
(\ref{Psimat}) is covariant under the  coaction (\ref{matbetaLR}).
\end{propos}
\proof We verify  covariance as
$(\vecs_1^{-1}\vecx'_1\vect_1)(\vecs_2^{-1}\vecx_2\vect_2)
=\vecs_1^{-1}\vecs_2^{-1}\vecx'_1\vecx_2\vect_1\vect_2
=\vecs_1^{-1}\vecs_2^{-1}R\vecx_2\vecx_1'R\vect_1\vect_2
=R\vecs^{-1}_2\vecs_1^{-1}\vecx_2\vecx'_1\vect_2\vect_1
R=R(\vecs_2^{-1}\vecx_2\vect_2)(\vecs_1^{-1}\vecx'_1\vect_1)R$
using the relations in each algebra and $[\vecs,\vect]=0$. Another way to see
this covariance is to write a multi-index R-matrix ${\bf
R}_+^I{}_J{}^K{}_L=R^{j_0}{}_{i_0}{}^{l_0}{}_{k_0}R^{i_1}{}_{j_1}
{}^{k_1}{}_{l_1}$ so that the braiding is the canonical form for a
braided covector space with $x_I=x^{i_0}{}_{i_1}$ in the framework
introduced in \cite{Ma:poi}. This is similar to the strategy used in
\cite{Ma:add} for addition of usual quantum matrices. One can easily
see that $A({\bf R}_+)$ with matrix generator $\Lambda^I{}_J$ maps onto
$A\tens A$ by $\Lambda^I{}_J=s^{-1}{}^{j_0}{}_{i_0}t^{i_1}{}_{j_1}$ and
its matrix right coaction $x_J\mapsto x_I\Lambda^I{}_J$ (with respect to
which the braiding is necessarily covariant) maps onto (\ref{matbetaLR}).
\endproof

Armed with this covariance, one has at once a $q$-Poincar\'e group according to
the general construction in \cite[Theorem~6]{Ma:poi}. This is generated by
$\vecp$ (a copy of $\vecx$ but regarded as momentum) and $\Lambda$ with cross
relations from ${\bf R}_+$. There is a dilaton element $\dila $ for
$\lambda_0\ne 1$. The spinorial version follows the same lines with the
rotations generated now by $\vecs,\vect$ and cross relations from $R$. The key
step is to note that $A\tens A$ has a tensor product dual-quasitriangular
structure consisting of $\CR$ on each factor. The coaction (\ref{matbetaLR})
becomes a right action of $A\tens A$ induced by evaluation against this as
\eqn{actpts}{ \vecp_1\ra\vect_2=  \vecp_1 \lambda_0 R,\quad
\vecp_1\ra\vecs_2=\lambda_0^{-1}R^{-1}\vecp_1.}
The semidirect product $\widetilde{(A\tens A)}\rcross \bar A(R)$ therefore has
the corresponding cross relations
\eqn{spinpoi}{ \vecp_1\vect_2=\lambda_0 \vect_2\vecp_1R,\quad
\vecp_1\vecs_2=\vecs_2\lambda_0^{-1} R^{-1}\vecp_1,\quad \vecp\dila
=\lambda_0^{-2}\dila \vecp.}
This computation follows the same steps as made recently for the q-Minkowski
case in \cite[Sec. 4]{MaMey:bra} but we see that the relations are
significantly simpler than there. The coproduct is the same, namely the
standard cross coproduct by the coaction of the form (\ref{matbetaLR}). As in
\cite{MaMey:bra} this comes out as $\Delta\vecp=\vecp\tens \vecs^{-1}(\
)\vect\dila +1\tens\vecp$ where the indices of $\vecp$ have to be inserted into
the space.

\section{ $q$-Euclidean space}

In this section we specialise to the standard R-matrix
\eqn{Rjones}{R=\pmatrix{q&0&0&0\cr 0&1&q-q^{-1}&0\cr 0&0&1&0\cr
0&0&0&q}}
and study $\bar A(R)$ constructed in the last section for this case. The usual
$A(R)$ is the usual quantum matrices $M_q(2)$.

The relations are computed from Proposition~3.2 for a matrix
$\vecx=\pmatrix{a&b\cr c&d}$ as
\cmath{ba=qab,\quad ca=q^{-1}ac,\quad da=ad,\qquad
bc=cb+(q-q^{-1})ad\\
db=q^{-1}bd\quad dc=qcd}
which we call the algebra of {\em $q$-Euclidean space} $\R_q^4=\bar M_q(2)$. We
know from the last section  that this algebra is fully covariant under the
action of $SU_q(2)\tens SU_q(2)$ by (\ref{matbetaLR}). This therefore plays the
role of the 4-dimensional rotation group in this picture. We note in passing
that $\bar M_q(2)\isom M_q(2)$ as an algebra by the permutation of generators
\eqn{isom}{ \pmatrix{a&b\cr c&d} \leftrightarrow \pmatrix{c&d\cr a&b}.}
Thus our construction does not give a genuinely new algebra in this simplest
case. On the other hand, our matrix coaction (\ref{matbetaLR}) and other of our
constructions would not appear very natural when mapped over to $M_q(2)$ by
this identification.

Next, it is easy to see that the standard $2\times 2$ quantum matrices $M_q(2)$
for real $q$ are a $*$-bialgebra
\eqn{eucl*}{\pmatrix{a^*&b^*\cr c^*&d^*}=\pmatrix{d&-q^{-1}c\cr -qb&a}}
and its dual-quasitriangular structure cf\cite{Ma:qua} is of real type. Hence
from Proposition~3.4 we know that our $q$-Euclidean space is also a $*$-algebra
with the operation (\ref{eucl*}).

Moreover, a short computation gives that the element $\und{det}=ad-q^{-1}bc$ is
central in the algebra $\R_q^4$ and, moreover self-adjoint. Thus we are
motivated to take it as a natural `metric'. If we make a change of variables
\eqn{xyzt}{t={a-d\over 2\imath},\quad x={c-qb\over 2}, \quad y={c+qb\over
2\imath},\quad z={a+d\over 2}}
then $t^*=t,x^*=x,y^*=y,z^*=z$ are some natural self-adjoint spacetime
co-ordinates and
\eqn{euclmetric}{ \und{det}=ad-q^{-1}bc=({1+q^{-2}\over
2})t^2+q^{-2}x^2+q^{-2}y^2+({1+q^{-2}\over 2})z^2}
so that the signature of the natural metric is the Euclidean one.

The multiplication law works with the multiplicative statistics (\ref{psimat})
between two copies, which come out as
\[a'a=q^{-1}aa'+(q^{-1}-q)bc',\quad a'b=ba',\quad
a'c=q^{-1}ca'+(q^{-1}-q)dc',\quad a'd=da'\]
\[b'a=q^{-1}ab'+(q^{-1}-q)bd',\quad b'b=bb',\quad
b'c=q^{-1}cb'+(q^{-1}-q)dd',\quad b'd=db'\]
\[c'a=ac',\quad c'b=q^{-1}bc',\quad c'c=cc',\quad c'd=q^{-1}dc'\]
\[d'a=ad',\quad d'b=q^{-1}bd',\quad d'c=cd',\quad d'd=q^{-1}dd'\]
times an overall factor $\lambda_0^{-1}=q^{\h}$ on the right hand sides, which
is not relevant for the multiplicativity property itself. The structure
however, is not a usual bialgebra, nor a braided one since the exchange law
$\psi$ underlying these statistics does {\em not} obey the QYBE.

Finally, there is a usual braided-Hopf algebra for addition with the additive
braid statistics (\ref{Psimat}). In our case this (like the algebra itself)
works out isomorphic under (\ref{isom}) to the additive braid statistics
already obtained for $M_q(2)$ in \cite{Ma:add}, so we do not repeat its listing
here. It means that our $q$-Euclidean 4-vectors, like $q$-Minkowski
4-vectors\cite{Mey:new}, can be added. Moreover, Proposition~3.5 assures us
that this addition is covariant under $SU_q(2)\tens SU_q(2)$ and that we have
an associated  4-dimensional Euclidean $q$-Poincar\'e group as a semidirect
product or bosonisation  $\widetilde{SU_q(2)\tens SU_q(2)}\rcross\R_q^4$ from
(\ref{spinpoi}). The multi-index generators $\Lambda^I{}_J$ there provide also
a vectorial picture of the Euclidean rotation group $O_q(4)$ in our framework
with corresponding $q$-Poincar\'e group $\widetilde{O_q(4)}\rcross \R_q^4$. In
both cases the scaling parameter is $\lambda_0=q^{-\h}\ne 1$ so a dilaton is
needed. In this way, we obtain the analogous structure to that found for
$q$-Minkowski space in \cite{Mey:new}\cite{MaMey:bra}.

\note{\[ a'a=q^2 aa',\quad b'b=q^2bb',\quad c'c=q^2 cc',\quad d'd=q^2 dd',\quad
a'b=qba'\]
\[ a'c=qca'+(q^2-1)ac',\quad a'd=da'+(q-q^{-1})bc',\quad b'a=qab'+(q^2-1)ba'\]
\[ b'c=cb'+(q-q^{-1})(da'+ad')+bc'(q-q^{-1})^2, \quad b'd=qdb'+(q^2-1)bd'\]
\[c'a=qac',\quad c'b=bc',\quad c'd=qdc'\]
\[ d'b=qbd',\quad d'c=qcd'+(q^2-1)dc',\quad  d'a=ad'+(q-q^{-1})bc'\]
which provides for the addition
\[ \pmatrix{a''&b''\cr c''& d''}=\pmatrix{a&b\cr c&d}+\pmatrix{a'&b'\cr
c'&d'}\]}

\section{$q$-Wick Rotation and Transmutation as Double-Twisting}

In this section we consider ourselves again in the general setting as in
Section~3 and give a new example and application of the twisting theory of
Section~2. Thus, consider $A$ a dual-quasitriangular Hopf algebra and this time
begin with the covariant system $(A\tens A,\bar A)$ obtained in
Proposition~3.3. This is our set-up for Euclidean space and its transformation
group, for example.

Consider now the cocycle on $A\tens A$ given by
\eqn{Lcocy}{ \chi((a\tens b)\tens (c\tens d))=\eps(a)\CR^{-1}(b\tens
c)\eps(d).}
One can easily see that it obeys the 2-cocycle condition (\ref{2-cocycle-on}).
In fact, it is nothing other than the dual of the cocycle $\CR_{23}^{-1}$ used
by \cite{ResSem:mat} in their construction of a`twisted square' Hopf algebra.
We have given the relevant dual construction for the Hopf algebra $A\dcross A$
in \cite[Sec. 4]{Ma:poi} as an algebra built on $A\tens A$ with the new product
\eqn{lorprod}{ (a\tens b)(c\tens d)= a c\t\tens b\t d\CR^{-1}(b\o\tens
c\o)\CR(b\th\tens c\th).}
We also gave on it a $*$-structure $(a\tens b)^*=b^*\tens a^*$ in the real case
and identified the spinorial $q$-Lorentz group of \cite{CWSSW:lor} as a matrix
example of this contruction. This abstract formulation of the $q$-Lorentz group
function algebra as the dual of the twisted square construction of
\cite{ResSem:mat} was one of the main results of the author in \cite{Ma:poi}.

\begin{propos} The covariant system $(A\tens A,\bar A)$ twists under the
cocycle (\ref{Lcocy}) to  the covariant system $(A\dcross A,\und A)$ where
$\und A$ has the product
\[ a\und\cdot b=a\t b\th \CR(a\th\tens Sb\o)\CR(a\o\tens b\t). \]
Here $A\dcross A$ is the double cross product Hopf algebra in \cite[Sec.
4]{Ma:poi} and $\und A$ is the braided group of function algebra type
introduced in \cite{Ma:bg}.
\end{propos}
\proof It is obvious that the twisted square in \cite{ResSem:mat} with
coproduct $\Delta_\chi=\CR_{23}^{-1}(\ )\CR_{23}$ is an example of twisting
with $\chi=\CR_{23}^{-1}$. The cocycle (\ref{Lcocy}) is just the dual of this
on our dual-quasitriangular Hopf algebra, and with the dual twisting it is
obvious that we have $(A\tens A)_\chi=A\dcross A$ where $A\dcross A$ has
product
(\ref{lorprod}). This can also be checked easily from (\ref{cotwistprod}). The
new part of the proposition concerns the twisting of $\bar A$. We compute from
(\ref{cotwistcomod}) its product twisted by (\ref{Lcocy}) as
\align{&&\equad a\cdot_\chi b=\chi^{-1}(Sa\o\tens a\th\tens Sb\o\tens b\th)
a\t\bar\cdot b\t\\
&&=\CR(a\t\tens Sb\o)a\o\bar\cdot b\t=\CR(a\th\tens Sb\o)\CR(a\o\tens b\t)a\t
b\th}
as stated.
\note{\align{&&\equad \beta(a)\beta(b)=a\t\cdot_\chi b\t\tens (Sa\o)(Sb\o)\tens
a\th b\th\\
&&=\CR(a\t\tens b\t)a\th b\th\tens S(b\o a\o)\tens a\fo b\fo\\
&&=\CR(a\o\tens b\o) a\th b\th\tens S(a\t b\t)\tens a\fo b\fo\\
&&=\CR(a\o\tens b\o) (a\t b\t)\t \tens S(a\t b\t)\o\tens (a\t b\t)\th}}
\endproof

The passage $A\mapsto \und A$ was introduced in one step in \cite{Ma:bg} as a
process of covariantisation or {\em transmutation} by means of a
category-theoretical (braided) Tannaka-Kein reconstruction theorem. Our new
mathematical result in this proposition is to factorise this complicated
construction into two steps: the construction of $\bar A$ followed by its {\em
Wick rotation} to $\und A$. Note also that  $A\dcross A$ maps onto $A$ by
multiplication. In this case the coaction (\ref{betaLR}) under which $\und A$
is covariant becomes the standard quantum adjoint coaction.

We call this twisting `Wick rotation' because for the standard R-matrix
(\ref{Rjones}) the algebra $SU_q(2)\dcross SU_q(2)$ is an established spinorial
form of the $q$-Lorentz group as explained above, while the algebra
$\und{M_q(2)}=BM_q(2)$ as given by transmuting the quantum matrices to braided
ones\cite{Ma:exa} is an established $q$-Minkowski space
\cite{Ma:mec}\cite{Ma:poi}\cite{Mey:new}\cite{MaMey:bra}
cf.\cite{CWSSW:ten}\cite{CWSSW:lor}\cite{OSWZ:def}. Thus our $q$-Euclidean
space and the action of $O_q(4)$ in the form $SU_q(2)\tens SU_q(2)$ gets
precisely `Wick rotated' by twisting to this established Minkowski space
approach. Since the Euclidean picture is important for the rigorous
construction of ordinary quantum field theory, it seems likely that our
$q$-Euclidean picture will be useful too in this programme of $q$-deforming
physics.

Finally, we note that there is an enveloping algebra version of all these
results based on quasitriangular Hopf algebras $H$. This time we use a cocycle
as in (\ref{2-cocycle}) to twist a coalgebra on which $H$ acts to a new
coproduct
\eqn{twimodcoalg}{ \Delta_\chi(c)=\chi\la\Delta c.}
The cocycle condition gives at once that this is coassociative. The twisted
quantum group $H_\chi$ acts on it covariantly in the sense that the action is a
coalgebra map (a module-coalgebra). In our specific application, we can take
$H$ as a coalgebra acted upon by $H\tens H^{\rm cop}$ according to
\eqn{HHact}{ (h\tens g)\la c=h c Sg}
which it the analogue of (\ref{betaLR}). Here $H^{\rm cop}$ means with the
opposite coalgebra. We take cocycle $\chi=\CR_{42}$ so that $(H\tens H^{\rm
cop})_\chi=H\tens H$. According to the general theory then, this acts
covariantly on the new coalgebra $\bar H$ with coproduct
\eqn{barH}{  \bar\Delta c=\CR_{42}\la\Delta c=(\Delta c)(S\tens
S)\CR_{21}=(\Delta c)\CR_{21}}
as $\CR$ is invariant under $(S\tens S)$. So far only the action of $H^{\rm
cop}$ was significant and we could obtain $\bar H$ as acted upon by $H$ equally
well, by starting with $H^{\rm cop}$ acting by $g\la c=cSg$ and cocycle
$\chi=\CR_{21}$, as the analogue of Proposition~3.1. Next we take this
covariant system $(H\tens H,\bar H)$ and twist again with cocycle
$\chi=\CR_{32}$. This gives a new coalgebra $\und H$ with coproduct
\eqn{undH}{ \und \Delta c= \CR_{32}\la\bar\Delta c=c\o \CR\ut S\CR'\ut \tens
\CR'\uo c\t \CR\uo=c\o S\CR\ut\tens \Ad_{\CR\uo}(c\t)}
where $\Ad$ is the quantum adjoint action. Here $\CR=\CR\uo\tens \CR\ut$
denotes the decomposition of $\CR$ as an element of $H\tens H$ and $\CR'$
denotes a second copy of it. We obtain exactly the transmuted coproduct $\und
H$ introduced in \cite{Ma:bra} and on it an action (\ref{HHact}) of the quantum
group $(H\tens H)_\chi$ which is a version of the twisted square of
\cite{ResSem:mat} with twisting by $\CR_{32}$. Its subalgebra $H$ also acts, by
the quantum adjoint action. This is of course nothing other than our results
above in a dualised form and a convenient left-right reversal. In principle, it
is not necessary to repeat the proofs again since there are diagrammatic
methods to make such dualisations (by turning the diagrams up-side-down).

In addition, there is nothing stopping the reader partly dualising our above
results, namely working with $\bar A$ or $q$-Euclidean space (not dualised) and
$H\tens H$ or $U_q(su_2)\tens U_q(su_2)$ acting on it. This dualisation is the
usual one and could be constructed using (\ref{twistmodalg}). The action from
(\ref{betaLR}) is $(h\tens g)\la a=\<Sh,a\o\> a\t\<g,a\th\>$ where $\<\ ,\ \>$
is the duality pairing. In the concrete matrix setting the action is
\eqn{lmact}{ l^+_1\la \vecx_2=\lambda_0^{-1}R^{-1}_{21}\vecx_2,\quad
l^-_1\la\vecx_2=\lambda_0R\vecx_2,\quad m^+_2\la \vecx_1=\vecx_1
\lambda_0R,\quad m^-_2\la\vecx_1=\vecx_1 \lambda_0^{-1}R^{-1}_{21}}
where $l^\pm$ and $m^\pm$ denotes the FRT generators\cite{FRT:lie} of the two
copies of $U_q(su_2)\tens U_q(su_2)$ or other quantum enveloping algebra. This
time the quantum Wick rotation proceeds by twisting further with cocycle
$\CR_{23}^{-1}$ and turns the $q$-Euclidean space $\bar M_q(2)$ to the
$q$-Minkowski space of braided-matrices and turns $U_q(su_2)\tens U_q(su_2)$ to
the twisted square $U_q(su_2)\codcross U_q(su_2)$ as the usual dual of the
spinorial $q$-Lorentz group function algebra $SU_q(2)\dcross SU_q(2)$. This
formulation of the $q$-Lorentz group enveloping algebra as a twisted square was
explained in \cite[Sec. 4]{Ma:poi} and subsequently reiterated by other authors
also. Moreover, since the $q$-Euclidean system is
covariant, our quantum Wick rotation ensures that
$q$-Minkowski space is likewise covariant as a module algebra under the
$q$-Lorentz group enveloping algebra, acting as in (\ref{lmact}) with $\vecx$
replaced by $\vecu$.

\subsection*{Acknowledgements} I would like to thank U. Meyer for ongoing
discussions on these topics.


\begin{thebibliography}{10}

\bibitem{Ma:exa}
S.~Majid.
\newblock Examples of braided groups and braided matrices.
\newblock {\em J. Math. Phys.}, 32:3246--3253, 1991.

\bibitem{Ma:mec}
S.~Majid.
\newblock The quantum double as quantum mechanics.
\newblock {\em J. Geom. Phys.}, 13:169--202, 1994.

\bibitem{CWSSW:ten}
U.~Carow-Watamura, M.~Schlieker, M.~Scholl, and S.~Watamura.
\newblock Tensor representation of the quantum group {$SL_q(2,\C)$} and quantum
  {M}inkowski space.
\newblock {\em Z. Phys. C}, 48:159, 1990.

\bibitem{CWSSW:lor}
U.~Carow-Watamura, M.~Schlieker, M.~Scholl, and S.~Watamura.
\newblock A quantum {L}orentz group.
\newblock {\em Int. J. Mod. Phys.}, 6:3081--3108, 1991.

\bibitem{OSWZ:def}
O.~Ogievetsky, W.B. Schmidke, J.~Wess, and B.~Zumino.
\newblock {$q$}-{D}eformed {P}oincar{\'e} algebra.
\newblock {\em Comm. Math. Phys.}, 150:495--518, 1992.

\bibitem{PodWor:def}
A.~Podles and S.L. Woronowicz.
\newblock Quantum deformation of {L}orentz group.
\newblock {\em Comm. Math. Phys}, 130:381--431, 1990.


\bibitem{Mey:new}
U.~Meyer.
\newblock A new {$q$}-{L}orentz group and $q$-{M}inkowski space with both
  braided coaddition and {$q$}-spinor decomposition.
\newblock {\em Preprint}, DAMTP/93-45, 1993.


\bibitem{Ma:poi}
S.~Majid.
\newblock Braided momentum in the {$q$}-{P}oincar{\'e} group.
\newblock {\em J. Math. Phys.}, 34:2045--2058, 1993.

\bibitem{MaMey:bra}
S.~Majid and U.~Meyer.
\newblock Braided matrix structure of {$q$}-{M}inkowski space and
  {$q$}-{P}oincar\'e group, december, 1993 (damtp/93-68).
\newblock {\em Z. Phys. C}.
\newblock To appear.


\bibitem{ResSem:mat}
N.Yu. Reshetikhin and M.A. Semenov-Tian-Shansky.
\newblock Quantum {$R$}-matrices and factorization problems.
\newblock {\em J. Geom. Phys.}, 5:533, 1988.

\bibitem{Dri:qua}
V.G. Drinfeld.
\newblock Quasi{H}opf algebras.
\newblock {\em Algebra i Analiz}, 1(6):2, 1989.
\newblock In Russian. Translation in {\em Leningrad Math J.}

\bibitem{Dri}
V.G. Drinfeld.
\newblock Quantum groups.
\newblock In A.~Gleason, editor, {\em Proceedings of the {ICM}}, pages
  798--820, Rhode Island, 1987. AMS.

\bibitem{FRT:lie}
L.D. Faddeev, N.Yu. Reshetikhin, and L.A. Takhtajan.
\newblock Quantization of {L}ie groups and {L}ie algebras.
\newblock {\em Leningrad Math J.}, 1:193--225, 1990.

\bibitem{Ma:lin}
S.~Majid.
\newblock Quantum and braided linear algebra.
\newblock {\em J. Math. Phys.}, 34:1176--1196, 1993.

\bibitem{Ma:bg}
S.~Majid.
\newblock Braided groups.
\newblock {\em J. Pure and Applied Algebra}, 86:187--221, 1993.

\bibitem{GurMa:bra}
D.I. Gurevich and S.~Majid.
\newblock Braided groups of {H}opf algebras obtained by twisting.
\newblock {\em Pac. J. Math.}, 162:27--44, 1994.

\bibitem{Ma:qua}
S.~Majid.
\newblock Quasitriangular {H}opf algebras and {Y}ang-{B}axter equations.
\newblock {\em Int. J. Modern Physics A}, 5(1):1--91, 1990.

\bibitem{Ma:clau}
S.~Majid.
\newblock Cross product quantization, nonabelian cohomology and twisting of
  {H}opf algebras.
\newblock In {\em Proc. Generalised Symmetries, Clausthal, Germany, July 1993}.
  World Sci.
\newblock To appear.

\bibitem{Kem:mul}
A.~Kempf.
\newblock Multiparameter {$R$}-matrices and generalised twisting methods.
\newblock In {\em Proc. XX DGM, New York, June, 1990}. World Sci.


\bibitem{Ma:add}
S.~Majid.
\newblock On the addition of quantum matrices, september 1993 (damtp/93-57).
\newblock {\em J. Math. Phys.}
\newblock To appear.


\bibitem{Ma:bra}
S.~Majid.
\newblock Braided groups and algebraic quantum field theories.
\newblock {\em Lett. Math. Phys.}, 22:167--176, 1991.


\end{thebibliography}
\end{document}